\documentclass[epj,referee]{svjour}

\usepackage{graphics}
\usepackage{dcolumn}
\usepackage{bm}

\newcommand{\la}{\left<}
\newcommand{\ra}{\right>}

\newcommand{\Nav}{\ensuremath{\la N \ra}}
\newcommand{\Npav}{\ensuremath{\la N^p \ra}}
\newcommand{\nonexp}{\ensuremath{K_p}}
\newcommand{\muchain}{\ensuremath{\mu_\mathrm{c}}}
\newcommand{\dmuchain}{\ensuremath{\delta\mu_\mathrm{c}}}

\newcommand{\rhochain}{\ensuremath{\rho_N}}
\newcommand{\pchain}{\ensuremath{P(N)}}
\newcommand{\ce}{\ensuremath{c_\mathrm{s}}}
\newcommand{\cep}{\ensuremath{c_{\mu}}}
\newcommand{\wep}{\ensuremath{w_p}}
\newcommand{\veff}{\ensuremath{\tilde{v}}}
\newcommand{\Ueff}{\ensuremath{u_n}}
\newcommand{\Utau}{\ensuremath{u_{t}}}
\newcommand{\Ustar}{\ensuremath{u^*}}

\begin{document}

\title{Non-extensivity of the chemical potential of polymer melts} 

\author{J.P. Wittmer\inst{1}, A. Johner\inst{1}, A. Cavallo\inst{1,2}, P. Beckrich\inst{1}, F. Crevel\inst{1},
 J. Baschnagel\inst{1}}
%
\institute{Institut Charles Sadron, 23 rue du Loess, BP 84047, 67034 Strasbourg Cedex 2, France
\and 
Dipartimento di Fisica, Universit\`a degli Studi di Salerno, I-84084 Fisciano, Italy
}
\date{\today}

\abstract{
%
Following Flory's ideality hypothesis the chemical potential of a test chain of length $n$ immersed into
a dense solution of chemically identical polymers of length distribution $\pchain$ 
is extensive in $n$. We argue that an additional contribution 
$\dmuchain(n) \sim +1/\rho\sqrt{n}$ arises 
($\rho$ being the monomer density)
for all $\pchain$ if $n \ll \Nav$ 
which can be traced back to the overall incompressibility of the solution leading 
to a long-range repulsion between monomers.
Focusing on Flory distributed melts we obtain 
$\dmuchain(n) \approx \left(1- 2 n/\Nav\right) / \rho \sqrt{n}$ for $n \ll \Nav^2$,
hence, $\dmuchain(n) \approx - 1/\rho \sqrt{n}$ if $n$ is similar to the typical 
length of the bath $\Nav$.
Similar results are obtained for monodisperse solutions.
Our perturbation calculations are checked numerically by analyzing the 
annealed length distribution $\pchain$ of linear equilibrium polymers generated by 
Monte Carlo simulation of the bond-fluctuation model. 
As predicted we find, e.g., the non-exponentiality parameter $\nonexp \equiv 1 - \Npav/p!\Nav^p$ 
to decay as $\nonexp \approx 1 / \sqrt{\Nav}$ for all moments $p$ of the distribution.
%
}

\authorrunning{Wittmer}
\titlerunning{Non-extensivity of the chemical potential of polymer melts}
\maketitle

%
\section{Introduction}
\label{sec_intro}

\begin{figure}[t]
\centerline{\resizebox{0.9\columnwidth}{!}{\includegraphics*{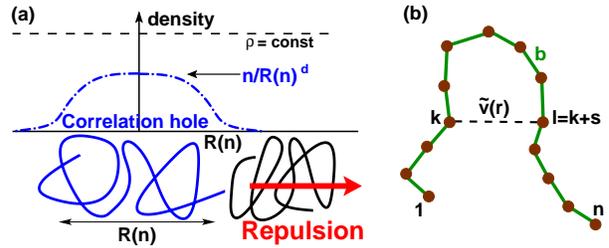}}}
\caption{
Sketch of the problem addressed:
{\bf (a)} 
Two chains repel each other in an incompressible melt since the total density $\rho$ cannot fluctuate. 
In $d$ dimensions this constraint leads to an entropic ``correlation hole potential
" $\Ustar \approx n/\rho R^d(n) \sim n^{1-d/2}$ one has to pay for joining both chains 
with $n$ and $R(n)$ being, respectively, the length and the typical size of the test chains
\cite{DegennesBook,ANS03}. This sets a non-extensive deviation 
$\dmuchain(n) \approx \Ustar(n)$ 
with respect to the chemical potential of asymptotically long chains.
{\bf (b)} We compute within one-loop perturbation theory the leading
non-extensive contribution to the mean self-interaction energy 
$\la \Ueff \ra_0$ of a test chain with respect to its Gaussian reference state,
$\Ueff$ being the sum of the effective monomer interactions $\veff(r)$ between pairs 
of monomers $k$ and $l=k+s$ at distance $r$.
\label{fig_sketch}
}
\end{figure}

%
%
One of the cornerstones of polymer physics is Flory's ideality hypothesis
\cite{DegennesBook,DoiEdwardsBook,SchaferBook}
which states that polymer chains in the melt follow Gaussian statistics,
i.e. they are random walks without long range correlations. 
The official justification of this mean-field result is that density fluctuations 
are small beyond the screening length $\xi$, hence, negligible \cite{DoiEdwardsBook}.
The size of a chain segment of arc-length $s$ of a test chain of length $n$
plugged into a melt of chemically identical $N$-polymers of (normalized) length distribution 
$\pchain$ and mean length $\Nav$ \cite{foot_PN} scales, hence, as
\begin{equation}
R^2(s) \equiv \la {\bm r}^2 \ra = b^2 s \mbox{ for } g \ll s \le n \ll \Nav^2
\label{eq_Rs_gauss}
\end{equation}
with $b$ denoting the effective bond length and
$g$ the number of monomers spanning the screening length $\xi$
\cite{DoiEdwardsBook}.
%
See Fig.~\ref{fig_sketch} for a sketch of some of the notations used in this paper.
%
%
The extensivity of the chemical potential $\muchain(n)$ of the test chain with respect to $n$, 
\begin{equation}
\muchain(n) = \mu n \mbox { for } g \ll n \ll \Nav^2,
\label{eq_muchain_gauss}
\end{equation}
$\mu > 0$ being the effective chemical potential {\em per} monomer,
is yet another well-known consequence of Flory's hypothesis \cite{DegennesBook}.
The upper boundary $\Nav^2$ indicated in Eq.~(\ref{eq_Rs_gauss}) 
and Eq.~(\ref{eq_muchain_gauss}) for later reference is due to the well-known swelling 
of extremely large test chains where the bath acts as a good solution 
\cite{DegennesBook,SchaferBook,Sergei81}.

Assuming Eq.~(\ref{eq_muchain_gauss}) to hold for the $N$-chains of the bath,
dense grand-ca\-no\-ni\-cal ``equilibrium polymers" \cite{CC90,WMC98b} are thus supposed to be 
``Flory size distributed", 
\begin{equation}
\pchain = \mu e^{-\mu N}, 
\label{eq_PN_gauss}
\end{equation}
where
as elsewhere in this paper, temperature and Boltzmann's constant have been set to unity
\cite{foot_FloryHuggins}.
Eq.~(\ref{eq_PN_gauss}) implies, of course, that $\Npav = p!/\mu^p$ for the $p$-th moment of the distribution.
Strictly speaking, Eq.~(\ref{eq_PN_gauss}) applies only for $g \ll N \ll \Nav^2$,
but both limits become irrelevant for systems of large mean chain length, $\mu \to 0$,
where only exponentially few chains are not within these bounds.

%
%
Recently, Flory's hypothesis has been challenged by the discovery of long-range
intrachain correlations in three dimensional melts 
\cite{Sergei81,WMBJOMMS04,WBJSOMB07,BJSOBW07,WBM07,papRouse,WCK09,papPrshort} 
and ultrathin films \cite{Sergei81,ANS03,CMWJB05}.
The physical mechanism of these correlations is related to the ``correlation hole" of density 
$n/R(n)^d$ in $d$ dimensions with $R(n)$ being the typical size of the test chain
(Fig.~\ref{fig_sketch}(a)). 
Due to the overall incompressibility of the melt this is known to set an entropic penalty 
$\Ustar(n) \approx n/\rho R(n)^d$ 
which has to be paid if two chains of length $n$ are joined together or which is
gained if a chain is broken into two parts \cite{ANS03}. The same effective repulsion 
acts also between adjacent chain segments of length $s$ and this on all scales 
\cite{WBJSOMB07,WBM07}.
In three dimensions this leads to a weak swelling of the chain segments characterized, e.g., by
\begin{equation}
1 - \frac{R^2(s)}{b^2 s} = \frac{\ce}{\sqrt{s}} 
\mbox{ for } g \ll s \ll n
\label{eq_ce_def}
\end{equation}
with $\ce = \sqrt{24/\pi^3} / \rho b^3$ denoting the ``swelling coefficient" 
\cite{WMBJOMMS04,WBCHCKM07,WBM07}.
Similar corrections with respect to Flory's ideality hypothesis have been obtained
for other intrachain {\em conformational} properties such as
higher moments of the segmental size distribution \cite{WBM07}, 
orientational bond correlations \cite{WMBJOMMS04,WBCHCKM07,papPrshort}
or the single chain structure factor \cite{WBJSOMB07,BJSOBW07}.
%

%
%
In this paper we question the validity of Flory's hypothesis for a central {\em thermodynamic} property, 
the chemical potential $\muchain(n)$ of a test chain inserted into a three-dimensional melt.
Our key claim is that the correlation hole potential leads to a deviation 
\begin{equation}
\dmuchain(n) \equiv \muchain - \mu n \approx \Ustar(n) \sim +1/\rho\sqrt{n}
\mbox{ for } n \ll \Nav
\label{eq_claim}
\end{equation}
that is non-extensive in chain length and this irrespective of the distribution $\pchain$ of the bath. 
Covering a broader $n$-range we will show explicitly for a melt of {\em quenched} Flory-size distribution
that
\begin{equation}
\dmuchain(n) \approx \frac{\cep}{\sqrt{n}} \left(1 - 2 \mu n \right)
\mbox{ for } g \ll n \ll \Nav^{2}
\label{eq_muEP}
\end{equation}
where we have defined $\cep = 3 \ce /8$.
This correction implies for the {\em annealed} length distribution of 
linear equilibrium polymers that (to leading order)
\begin{eqnarray}
\pchain & \approx&  \mu e^{-\mu N - \dmuchain(N)}
\label{eq_PNclaim1}
\\
&\approx& \mu e^{-\mu N} \left( 1 - \frac{\cep}{\sqrt{N}} (1-2\mu N) \right)
\label{eq_PNclaim}
\end{eqnarray}
where both the lower ($g \ll N$) and the upper limit ($N \ll \Nav^2$) of validity
become again irrelevant in the limit of large mean chain length.
Eq.~(\ref{eq_PNclaim}) will allow us to demonstrate Eq.~(\ref{eq_muEP}) numerically from the observed 
non-ex\-ponen\-tiality 
of the length distribution of equilibrium polymer melts obtained by means of Monte Carlo simulation
of a standard lattice model \cite{WMC98b,BFM}.

%
%
The one-loop perturbation calculation leading to Eq.~(\ref{eq_muEP}) is presented in Section~\ref{sec_theo}
where we will also address the chemical potential of monodisperse melts. Section~\ref{sec_algo} outlines the numerical algorithm 
used for the simulation of equilibrium polymers. 
Our computational results are compared 
to theory in Section~\ref{sec_simu}.

\section{Perturbation calculation}
\label{sec_theo}

\subsection{General remarks}
\label{sub_theo_general}
Following Edwards~\cite{DoiEdwardsBook} we take as a reference for the perturbation calculation 
a melt of Gaussian chains of effective bond length $b$. Averages performed over this unperturbed 
reference system are labeled by an index $0$. The general task is to compute the ratio $Q(n)/Q_0(n)$
of the perturbed to the unperturbed single chain partition function
\begin{eqnarray}
1 - \frac{Q(n)}{Q_{0}(n)}  
& = & 1- \la e^{-\Ueff} \ra_0 \approx
\nonumber \label{eq_task1} \\
\la \Ueff \ra_0 
& = & \sum_{s=0}^n (n-s) 
\int d{\bm r} \ G(r,s) \veff(r)
\label{eq_task2}
\end{eqnarray}
with the perturbation potential $\Ueff$ being the sum of the effective monomer interactions 
$\veff(r)$ of all pairs of monomers of the test chain of length $n$,
and $G(r,s)$ denoting the Gaussian propagator for a chain segment of length $s$ \cite{DoiEdwardsBook}. 
The factor $n-s$ in Eq.~(\ref{eq_task2}) counts the number of equivalent monomer pairs
separated by an arc-length $s$.
The deviation $\dmuchain(n)$ from Flory's hypothesis is then
given by the contribution to $\la \Ueff \ra_0$ which is non-linear in $n$.
The calculation of Eq.~(\ref{eq_task2}) in $d$ dimensions is most readily performed 
in Fourier-Laplace space with $q$ being  the wavevector conjugated to the monomer distance $r$
and $t$ the Laplace variable conjugated to the chain length $n$.
The Laplace transformed averaged perturbation potential reads
\begin{eqnarray}
\Utau & \equiv & \int_{n=0}^{\infty} dn \la \Ueff \ra_0 e^{-n t} 
\\
& = &
\int \frac{d^dq}{(2\pi)^d} \frac{1}{t^2} G(q,t) \veff(q)
\label{eq_task3}
\end{eqnarray}
where the factor $1/t^2$ accounts for the combinatorics and
$G(q,t)= 1/((aq)^2+t)$ represents the Fourier-Laplace transformed Gaussian propagator $G(r,s)$
\cite{DoiEdwardsBook,papPrshort}
with $a \equiv b/\sqrt{2d}$ being a convenient monomeric length.

\subsection{Effective interaction potential}
\label{sub_theo_veff}

We have still to specify $\veff(q)$, the effective interaction between test chain 
monomers in reciprocal space. This interaction is partially screened by the background 
of the monomers of the bath. It has been shown by Edwards \cite{DoiEdwardsBook,BJSOBW07} that
within linear response this corresponds to
\begin{equation}
\frac{1}{\veff(q)} = \frac{1}{v} + F_0(q) \rho.
\label{eq_veff_def}
\end{equation}
The bare excluded volume $v$ indicated in the first term of Eq.~(\ref{eq_veff_def}) characterizes 
the short-range repulsion between the monomers. Thermodynamic consistency requires
\cite{ANS05a,BJSOBW07,WCK09} that $v$ is proportional to the inverse of the
measured compressibility of the solution
\begin{equation}
v =  \frac{1}{g\rho} \equiv \frac{1}{2\rho} \ (a/\xi)^2
\label{eq_vbare}
\end{equation}
where we have defined the screening length $\xi$ following, e.g.,
eq~5.38 of Ref.~\cite{DoiEdwardsBook}.
Interestingly, the number $g$ of monomers spanning the blob, i.e., the lower bound of validity
of various statements made in the Introduction, can be determined experimentally 
or in a computer simulation from the low-wavevector limit of the total monomer structure factor and, 
due to this operational definition, $g$ is sometimes called ``dimensionless compressibility" 
\cite{WCK09,papPrshort}.
$F_0(q)$ stands for the ideal chain intramolecular structure factor of the given distribution
$\pchain$ of the bath. The effective interaction, Eq.~(\ref{eq_veff_def}), depends thus in 
general on the length distribution of the melt the test chain is inserted.
For Flory distributed melts the structure factor is, e.g., given by \cite{BJSOBW07}
\begin{equation}
F_0(q) = \frac{2}{(a q)^2 + \mu}
\label{eq_Fq_FD}
\end{equation}
while for monodisperse melts it reads $F_0(q) = N f_D(y)$ with $y= N (a q)^2$ and
$f_D(y) = 2(e^{-y}-1+y)/y^2$ being Debye's function \cite{DoiEdwardsBook}.
We remind that within the Pad\'e approximation for monodisperse chains Eq.~(\ref{eq_Fq_FD}) 
holds with $\mu$ replaced by $2/N$ \cite{DoiEdwardsBook}.
%

Below we will focus on incompressible solutions ($g=1/v\rho\to 0$) 
where the effective interaction is given by
the inverse structure factor, $\veff(q) \approx 1/\rho F_0(q)$, i.e. we will ignore local
physics on scales smaller than the correlation length $\xi$ and assume that both the
test chain and the chains of the bath are larger than $g$.
The effective potential takes simple forms at low and high wavevectors
corresponding, respectively, to distances much larger or much smaller than
the typical size of the bath chains.
In the low-wavevector limit the potential becomes 
$\veff_0 \equiv \veff(q\to 0) = \la N \ra / \la N^2 \ra \rho$ for general $\pchain$, 
i.e. $\veff_0 = 2\mu/\rho$ for Flory distributed and $\veff_0 = 1/\rho N$ for monodisperse melts. 
Long test chains are ruled by $\veff_0$ which acts as a weak repulsive pseudo-potential with 
associated (bare) Fixman parameter $z = \veff_0 \sqrt{n}$. As already recalled in the Introduction,
test chains with $n \gg (\Nav^2/\Nav)^2 \approx \Nav^2$ \cite{foot_PN} must thus swell 
and obey excluded volume statistics \cite{DegennesBook,SchaferBook}.

The effective potential of incompressible melts becomes scale free for larger wavevectors 
corresponding to the self-similar random walks,
\begin{equation}
\veff(q) \approx \frac{(a q)^2}{2\rho}
\mbox{ for } 1/q \ll b \Nav^{1/2},
\label{eq_veff_q2}
\end{equation}
i.e. the interactions decrease as a power law with distance and this irrespective of the
length distribution $\pchain$. One expects that short test chains with $n \ll \Nav$ see an 
interaction potential of effectively infinite bath chains as described by Eq.~(\ref{eq_veff_q2}).
Please note that Eq.~(\ref{eq_veff_q2}) lies at the heart of the power-law swelling of chain segments, 
Eq.~(\ref{eq_ce_def}), and related properties alluded to above \cite{WMBJOMMS04,BJSOBW07,WBM07,papPrshort}. 

\subsection{Ultraviolet divergency}
\label{sub_theo_regular}

Coming back to the computation of Eq.~(\ref{eq_task3}) one realizes that a naive
perturbation calculation using the effective interaction given in Eq.~(\ref{eq_veff_def})
is formally diverging at high wavevectors in three dimensions 
(becoming only regular below $d=2$) due to the monomer self-interactions
which should be subtracted. 
Using Eq.~(\ref{eq_veff_q2}) instead of Eq.~(\ref{eq_veff_def}) even makes things worse
due to an additional divergency associated with the self-interactions of the blobs
whose size was set to zero ($g\to 0$). However, since we are not interested in 
(possibly diverging) contributions linear in the length of the test chain or independent of it,
we can freely subtract linear terms (i.e., terms $\sim 1/t^2$ in Laplace space) 
or constant terms (i.e., terms $\sim 1/t$) to regularize and to simplify $\Utau$. 
Such a transformation leads to
\begin{equation}
\Utau =
\int \frac{d^dq}{(2\pi)^d} \frac{1}{t^2} \frac{2 F_0^{-1}-(a q)^2 - t}{(a q)^2+t}
\frac{v}{2(v\rho+F_0^{-1})} + \ldots
\label{eq_task4}
\end{equation}
where ``$\ldots$" stands for the linear and constant contributions we do not compute.
Converging now for incompressible melts for $d < 2$, the latter reformulation will prove useful below.
(See Section~\ref{sub_theo_d3} for the complete regulization of the ultraviolet divergency 
for incompressible three dimensional melts.)

\begin{figure}
\centerline{\resizebox{0.90\columnwidth}{!}{\includegraphics*{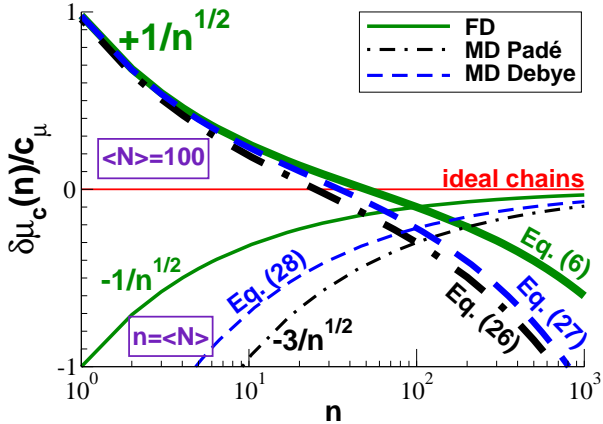}}}
\caption{
Non-extensive deviation of the chemical potential $\dmuchain(n)/\cep$ as a function of the test chain
length $n$. Flory distributed (FD) melts are indicated by solid lines,  monodisperse (MD) melts 
by dash-dotted and dashed lines for Pad\'e approximation and full Debye function, respectively.
The bold lines refer to test chains in melts of constant (mean) chain length with $\Nav=100$.
The deviation changes sign at $n \approx \Nav$.
The thinner lines refer to test chains of same length as the typical melt chain, $n \equiv \Nav$,
where the asymptotic Gaussian behavior ($\dmuchain(n)=0$) is approached systematically from below. 
\label{fig_muchain}
}
\end{figure}

\subsection{Flory-distributed melts}
\label{sub_theo_FD}
Applying Eq.~(\ref{eq_task4}) to incompressible Flory distributed melts this leads to 
\begin{eqnarray}
\Utau & =& \frac{1}{2\rho} \frac{\mu-t}{t^2}
\int \frac{d^dq}{(2\pi)^d} G(q,t)
+ \ldots 
\label{eq_FD1} \\
& = & \frac{1}{2\rho} (\mu/t^2 -1/t) G(r=0,t) + \ldots
\label{eq_FD1b}
\end{eqnarray}
where we have read Eq.~(\ref{eq_FD1}) as an inverse Fourier transform taken at $r=0$.
Remembering that a factor $1/t$ in $t$-space stands for an integral $\int_0^n \rm{d}s$ in $n$-space,
the inverse Laplace transform of $\Utau$ can be expressed in terms of integrals of the return probability 
$G(r=0,s) = (4\pi s a^2)^{-d/2}$. We obtain, hence, in $n$-space
\begin{equation}
\dmuchain(n) = 
\frac{1}{(d-2)(4\pi)^{d/2}}\frac{1}{ \rho a^d}\left(n^{1-d/2} - \mu \frac{ n^{2-d/2}}{2-d/2}\right)
\label{eq_FD2}
\end{equation}
where $\dmuchain(n)$ stands for the non-extensive contribution to $\la \Ueff \ra_0$.
Note that the first term in the brackets scales as the correlation hole in $d$ dimension.
Its marginal dimension is $d=2$. The second term characterizes the effective two-body interaction
of the test chain with itself. As one expects \cite{DegennesBook}, its marginal dimension is $d=4$.
Although Eq.~(\ref{eq_FD2}) is formally obtained for $d<2$ it applies to higher dimensions 
by analytic continuation. 

In three dimensions Eq.~(\ref{eq_FD2}) becomes
\begin{equation}
\dmuchain(n) = \frac{1}{(4\pi)^{3/2}}\frac{1}{ \rho a^3} \left(n^{-1/2} - 2\mu n^{1/2}\right)
\label{eq_FD3}
\end{equation}
which demonstrates finally the non-extensive correction to the ideal polymer chain 
chemical potential announced in Eq.~(\ref{eq_muEP}) and sketched for $1/\mu=\Nav=100$ 
in Fig.~\ref{fig_muchain} by the bold solid line. 
(A slightly different demonstration is given below in Section~\ref{sub_theo_d3}.) 
As anticipated above, the first term in Eq.~(\ref{eq_FD3}) dominates for short test chains.
It is independent of the polydispersity and scales as the correlation hole potential,
Eq.~(\ref{eq_claim}). The second term dominates for large test chains with $n > 1/(2\mu)$
becoming non-perturbative for $n \gg 1/\mu^2$. Please note that these extremely long chains
are essentially absent in the Flory bath but may be introduced on purpose. 
Both contributions to $\dmuchain(n)$ {\em decrease} with increasing $n$. They correspond to an effective
enhancement factor of the partition function quite similar to the $\dmuchain(n) = -(\gamma-1) \log(n)$ 
in the standard excluded volume statistics with $\gamma \approx 1.16 > 1$ being the self-avoiding walk 
susceptibility exponent \cite{DegennesBook}. 

Interestingly, while $\dmuchain(n)$ decreases at fixed $\Nav$ it {\em increases} as
$\dmuchain(n) = -\cep/\sqrt{n}$ for a test chain with $n \equiv \Nav$, 
as shown by the thin solid line in Fig.~\ref{fig_muchain}.  
The chemical potential of typical chains of the bath approaches thus the
Gaussian limit from below.

\subsection{Equilibrium polymers}
\label{sub_theo_EP}
Flory distributed polymer melts are obtained naturally in systems of self-assembled linear 
equilibrium polymers where branching and the formation of closed rings are forbidden
\cite{CC90,WMC98b,foot_FloryHuggins}. Since the suggested correction $\dmuchain(N)$ to the 
ideal chain chemical potential is weak the system must remain to leading order Flory distributed
and Eq.~(\ref{eq_muEP}) should thus hold \cite{foot_annealed}.
Using $P(N)/P_0(N) = Q(N)/Q_0(N)$ one obtains directly the corrected length distribution 
for equilibrium polymers announced in Eq.~(\ref{eq_PNclaim}).
Note that Eq.~(\ref{eq_PNclaim}) is properly normalized,
i.e. the prefactor $\mu$ of the distribution remains exact
if $\dmuchain(N)$ is given by Eq.~(\ref{eq_muEP}). 
Since the distribution becomes broader the first moment increases slightly at given $\mu$:
\begin{equation}
\Nav = \mu^{-1} \left(1+ \cep \sqrt{\mu \pi} \right).
\label{eq_mu2Nav}
\end{equation}
More generally, one expects for the  $p$th moment 
\begin{equation}
\Npav = \frac{p!}{\mu^p}
\left( 1 + \frac{\cep \sqrt{\mu}}{p!} \left( 2\Gamma(p+3/2) - \Gamma(p+1/2)\right)\right)
\label{eq_pmoment}
\end{equation}
with $\Gamma(x)$ being the Gamma function \cite{abramowitz}.
The non-ex\-po\-nen\-tial\-ity parameter $\nonexp \equiv 1 - \Npav / p! \Nav^p$ should thus scale as
\begin{equation}
\nonexp = \wep \cep \sqrt{\mu}
\label{eq_nonexp}
\end{equation}
with $\wep \equiv (\Gamma(p+1/2) + \sqrt{\pi} p p! -2 \Gamma(p+3/2))/p!$
being a $p$-dependent geometrical factor.
Eq.~(\ref{eq_nonexp}) will be tested numerically in Section~\ref{sec_simu}.

\subsection{Incompressible melts in three dimensions}
\label{sub_theo_d3}

It is instructive to recover Eq.~(\ref{eq_FD3}) directly in three dimensions. 
For that purpose we may subtract $1/(a q)^2$ from the propagator $G(q,t)$ in Eq.~(\ref{eq_task4})
which amounts to take off a linear and a constant contribution. Taking the incompressible
limit this yields 
\begin{equation}
\Utau = \int \frac{d^3 q}{(2\pi)^3}\frac{1}{t^2}\frac{2/F_0(q) - (a q)^2 - t}{t + (a q)^2}\frac{-t}{(a q)^2}\frac{1}{2\rho} + \ldots 
\label{eq_reg3d}
\end{equation}
for a general structure factor $F_0(q)$.
Assuming Eq.~(\ref{eq_Fq_FD}) for $F_0(q)$ we obtain by straightforward integration over momentum 
$\Utau = -(8\pi \rho a^3)^{-1}(\mu-t)/t^{3/2} + \ldots$. After taking the inverse Laplace transform this
confirms Eq.~(\ref{eq_FD3}).
Interestingly, the inverse Laplace transform of the general formula, Eq.~(\ref{eq_reg3d}), 
can be performed leading to
\begin{eqnarray}
\dmuchain(n) & = & \frac{1}{2\rho} \int \frac{d^3q}{(2\pi)^3} \left(\exp(-n (a q)^2) - f_c(q,n) \right) 
\nonumber
\\
& = & \frac{\cep}{\sqrt{n}} - \frac{1}{2\rho} \int \frac{d^3q}{(2\pi)^3} f_c(q,n)
\label{eq_reg3dRSb}
\end{eqnarray}
where
$f_c(q,n) \equiv \left(2 \Nav/F_0(q) - y\right) (1 - \exp(- x y))/y$
with $x \equiv n/\Nav$ and $y \equiv \Nav (a q)^2$.
The first term in Eq.~(\ref{eq_reg3dRSb}) corresponds to the infinite bath chain limit ($x \to 0$)
which does not depend on the length distribution $\pchain$.
The integral over $f_c(q,n)$ stands for finite-$x$ corrections
for larger test chains.
%

\subsection{Monodisperse melts}
\label{sub_theo_MD}

We turn now to incompressible monodisperse melts in three dimensions.
As already mentioned above, the Debye function for monodisperse melts
can be approximated by the structure factor of Flory distributed melts,
Eq.~(\ref{eq_Fq_FD}), replacing $\mu$ by $2/N$. It follows thus
from Eq.~(\ref{eq_muEP}) that within Pad\'e approximation we expect
\begin{equation}
\dmuchain(n) \approx \frac{\cep}{\sqrt{n}} \left(1 - 4 x \right)
\label{eq_mu_MD1}
\end{equation}
for $n \ll N^{2}$ with $x=n/N$. This is indicated by the dash-dotted line in Fig.~\ref{fig_muchain}.
If the test chain and the bath chains are of equal length,  $x=1$,
this leads to $\dmuchain(n) = - 3 \cep /\sqrt{n}$,
i.e. $\muchain(n)$ approaches again its asymptotic limit from below (thin dash-dotted line).

The calculation of the chemical potential deviations for the full Debye function
can be performed taking advantage of Eq.~(\ref{eq_reg3dRSb}). Specializing the formula to the
monodisperse case this yields after some simple transformations
\begin{equation}
\dmuchain(n) =  
\frac{\cep}{\sqrt{n}} \left(1 - \ I_c(x) \right)
\label{eq_mu_MD_Debye}
\end{equation}
where the finite-$x$ correction is expressed by the integral
$$I_c(x) \equiv
\sqrt{\frac{x}{\pi}}
\int_0^\infty{\underline{\left(\frac{2}{{\rm f}_D(y)} - y\right)} 
\left(\frac{1-\exp(-x y)}{y} \right) \frac{dy}{\sqrt y}}.$$
Eq.~(\ref{eq_mu_MD_Debye}) can be evaluated numerically as shown in Fig.~\ref{fig_muchain}
where the bold dashed line corresponds to a variation of $n$ at constant $N=100$
and the thin dashed line to a test chain of same length as the chains of the bath, $x=1$.
Both lines are bounded by the predictions for Flory distributed melts and
the Pad\'e approximation of monodisperse chains.

The evaluation of Eq.~(\ref{eq_mu_MD_Debye}) deserves some comments.
%
The underlined bracket under the integral $I_c(x)$ defines a slowly varying function of $y$ 
decreasing from $2$ to $1$ when $y$ increases from $0$ to infinity.
Without this slow factor the integral can be scaled: 
It is proportional to $x$ and evaluates to $I_c(x) = 2x$
in agreement with the correction term obtained for Flory distributed polymers, Eq.~(\ref{eq_muEP}). 
The integral is mostly build up by the region $x y<A$ with an error $\sim 1/\sqrt A$. 
For large $x \gg 1$ only small $y$ contribute to the integral;
the underlined term in the integral can be replaced by $2$ and we obtain
asymptotically $I_c(x) = 4 x$ in agreement with the Pad\'e approximation, Eq.~(\ref{eq_mu_MD1}). 
If the first subdominant contribution to the integral is also computed
one gets $I_c(x) =  4 x - 1.13 \sqrt{x}$ for large $x$.
For small test chains, the integral provides the first correction $I_c(x) = 2 x$,
i.e., it vanishes for $x\rightarrow 0$ as already noted. 
In short, we recover the known asymptotic for short and long test chains but the crossover is very sluggish. 
The simple Pad\'e approximation, Eq.~(\ref{eq_mu_MD1}),   
is off by $\approx 40\%$ in the crossover region where $x \approx 1$.
Note finally that if the test chain is a chain of the bath 
($x=1$) one evaluates numerically $I_c(x=1) = 3.19$. 
We obtain thus
\begin{equation}
\dmuchain(n) = - 2.19 \cep/\sqrt{n} 
\label{eq_mu_MD_Debye2}
\end{equation}
as indicated by the thin dashed line.

\section{Algorithmic issues}
\label{sec_algo}

%
%
The theoretical predictions derived above should hold in any sufficiently dense polymer 
solution assuming that the chains are not too short.
Since the direct measurement of the chemical potential of mono\-disperse chains
(discussed in Section~\ref{sub_theo_MD}) requires a delicate thermodynamic integration 
\cite{FrenkelSmitBook,MP94,WCK09}
we test the theoretical framework by computing numerically the 
length distribution $\pchain$ in systems of {\em annealed} equilibrium polymers 
\cite{foot_annealed}. 
The presented configuration ensembles have been obtained 
using the well-known ``bond fluctuation model" (BFM)
\cite{BFM,Deutsch,WMC98b}
--- an efficient lattice Monte Carlo scheme where a coarse-grained mono\-mer
occupies 8 lattice sites on a simple cubic lattice
(i.e., the volume fraction is $8\rho$)
and bonds between monomers can vary in length and direction.
%
All length scales are given in units of the lattice constant.
Systems with an annealed size distribution are obtained
by attributing a finite scission energy $E$ to each bond
which has to be paid whenever the bond between two monomers is broken.
Standard Metropolis Monte Carlo is used to reversibly break and recombine the chains 
\cite{WMC98b,HXCWR06}.
Branching and formation of closed rings are forbidden.
Only local hopping moves have been used since the breaking and recombination of chains 
reduce the relaxation times dramatically compared to monodisperse systems \cite{HXCWR06}.

%
%
We only present data for one high density where half of the lattice sites are occupied 
($\rho = 0.5/8$). It has been shown \cite{WBM07,WCK09} that for this density 
we have a dimensionless compressibility $g = 0.24$, i.e. the system may be regarded as
incompressible on all scales and the lower bound of validity of the theory is irrelevant,
and a swelling coefficient $\ce = 0.41$.
Hence, $\cep = 3 \ce/8 \approx 0.16$ for the only parameter of the theory tested here. 
We use periodic simulation boxes of linear length $L=256$ containing 
$2^{20} \approx 10^6$ monomers.
The scission energy $E$ has been increased systematically up to $E=15$ 
which corresponds to a mean chain length $\Nav\approx 6100$. 
The configurations used here have already been tested and analyzed in previous publications
discussing the non-ideal behavior of configurational intrachain properties 
\cite{WBCHCKM07,WBJSOMB07,BJSOBW07,papPrshort}.

\section{Computational results.}
\label{sec_simu}

\begin{figure}[t]
\centerline{\resizebox{0.9\columnwidth}{!}{\includegraphics*{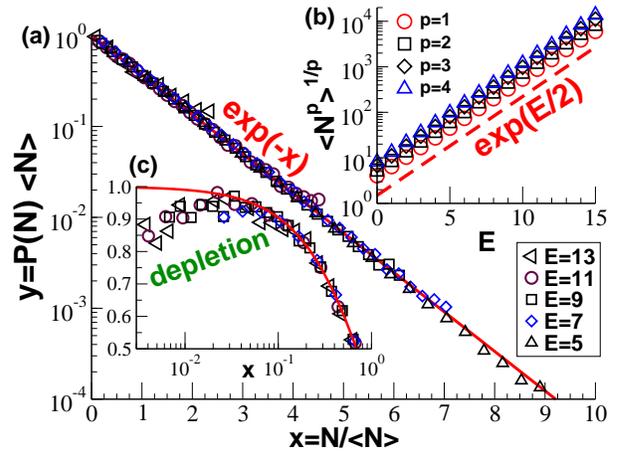}}}
\caption{
Chain length distribution $\pchain$ of linear equilibrium polymers 
for different scission energies $E$ obtained at number density $\rho=0.5/8$ of the BFM
Monte Carlo algorithm:
{\bf (a)} The main panel shows the collapse of the rescaled distribution $y=\pchain \Nav$ 
as a function of $x=N/\Nav$ for several $E$ as indicated. The exponential decay (solid line)
implied by Flory's ideality hypothesis is apparently (to first order) consistent with our data.
{\bf (b)} First four moments of the distribution {\em vs.} $E$.
{\bf (c)} Replot of the data of panel (a) in log-linear coordinates focusing on short chains.
The data points are systematically {\em below} the exponential decay (solid line) for $x \ll 1$.
\label{fig_Florydistr}
}
\end{figure}

%
The main panel of Figure~\ref{fig_Florydistr} presents the normalized length distribution $\pchain$
for different scission energies $E$ as indicated. A nice data collapse is apparently
obtained if $\pchain \Nav$ is plotted as a function of the reduced chain length $x =N/\Nav$
using the measured mean chain length $\Nav$. At first sight, there is {\em no} sign of
deviation from the exponential decay indicated by the solid line.
The mean chain length itself is given in panel (b) as a function of $E$ together with
some higher moments $\Npav= \sum_N N^p \pchain$ of the distribution.
As indicated by the dash\-ed line, we find $\Npav^{1/p} \sim \exp(E/2)$ as expected 
from standard linear aggregation theory \cite{CC90,WMC98b,foot_FloryHuggins}. 
The data presented in the first two panels of Fig.~\ref{fig_Florydistr} is thus fully consistent 
with older computational work \cite{WMC98b,HXCWR06} which has let us to believe that
Flory's ideality hypothesis holds rigorously. 

Closer inspection of the histograms reveals, however, deviations for small $x \ll 1$.
As can be seen from panel (c), the probability for short chains is reduced with
respect to the Flory distribution indicated by the solid line. 
This depletion agrees, at least qualitatively, 
with the predicted positive deviation of the chemical potential, Eq.~(\ref{eq_claim}).
 
\begin{figure}[t]
\centerline{\resizebox{0.9\columnwidth}{!}{\includegraphics*{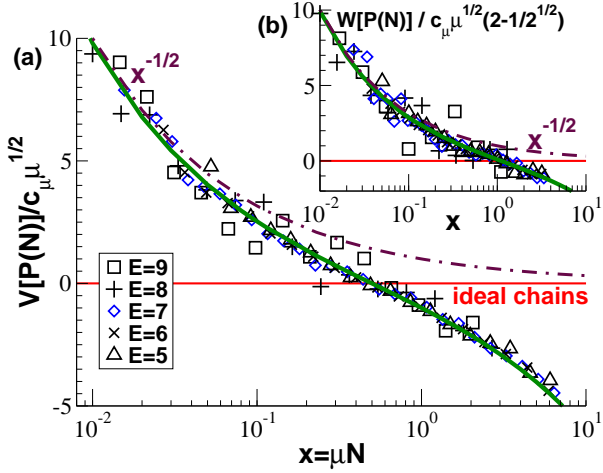}}}
\caption{
Characterization of the deviations of $\pchain$ from Eq.~(\ref{eq_PN_gauss}) 
using the functionals 
{\bf (a)} $V[P(N)]  \approx \dmuchain(N)$ and 
{\bf (b)} $W[P(N)] \approx 2 \dmuchain(N) - \dmuchain(2N)$ 
defined in the main text
which should both vanish for perfectly Flory distributed ideal chains as indicated by
the horizontal lines in the panels.
Data points for different scission energies $E$ (as indicated) collapse 
if $V[P(N)]/\cep \sqrt{\mu}$ and $W[P(N)]/\cep \sqrt{\mu} (2-1/\sqrt{2})$ 
are plotted as functions of the reduced chain length $x=\mu N$. 
For small $x$ both functionals decay as $1/\sqrt{x}$ as shown by the dash-dotted lines.
The bold lines correspond to the full predictions Eq.~(\ref{eq_VN}) and Eq.~(\ref{eq_WN})
for $V[P(N)]$ and $W[P(N)]$, respectively. 
\label{fig_WN}
}
\end{figure}

%
The curvature of $-\log(\pchain)$, i.e. the non-extensive deviation of the chemical potential
from Flory's ideality hypothesis, is further analyzed in Figure~\ref{fig_WN}.
Motivated by Eq.~(\ref{eq_PNclaim1}), we present in panel (a) the functional
\begin{equation}
V[P(N)] \equiv - \log(P(N)) - \mu N + \log(\mu)
\label{eq_VN_def}
\end{equation}
where the second term takes off the ideal contribution to the chemical potential.
The last term is due to the normalization of $\pchain$ and eliminates a 
trivial vertical shift depending on the scission energy $E$.
Consistently with Eq.~(\ref{eq_mu2Nav}), the chemical potential per monomer $\mu$ has
been obtained from the measured mean chain length $\Nav$ using
\begin{equation}
\mu \equiv {\Nav}^{-1} \left(1+ \cep \sqrt{\pi} / \sqrt{\Nav} \right).
\label{eq_Nav2mu}
\end{equation}
Note that $\mu$ and $1/\Nav$ become numerically indis\-tin\-guish\-able for $E \ge 7$. 
If the Gaussian contribution to the chemical potential is properly subtracted
one expects to obtain directly the non-Gaussian deviation to the chemical potential, 
$\dmuchain(N) \approx V[P(N)]$.  
Due to Eq.~(\ref{eq_muEP}) the functional should thus scale as
\begin{equation}
V[P(N)] / \cep \sqrt{\mu} \approx (1 - 2 x)/\sqrt{x}
\label{eq_VN}
\end{equation}
with $x = \mu N$ as indicated by the bold line in the panel. 
This is well born out by the data collapse obtained up to $x \approx 5$.
Obviously, the statistics detoriates for $x \gg 1$ for all energies 
due to the exponential cut-off of $\pchain$.
Unfortunately, the statistics of the length histograms decreases strongly with $E$
and becomes too low for a meaningful comparison for $E > 9$.
It is essentially for this numerical reason that we use Eq.~(\ref{eq_Nav2mu}) rather 
than simple large-$E$ limit $\mu = 1/\Nav$ since this allows us to add the two histograms 
for $E=5$ and $E=6$ for which high precision data is available. Otherwise these 
energies would deviate from Eq.~(\ref{eq_VN}) for large $x$ due to an insufficient
substraction of the leading Gaussian contribution to the chemical potential.
Thus we have used to some extend in panel (a) the predicted behavior, 
Eq.~(\ref{eq_muEP}), presenting strictly speaking a (highly non-trivial)
self-consistency check of the theory.

Since the substraction of the large linear Gaussian contribution is in any case
a delicate issue we present in panel (b) of Fig.~\ref{fig_WN} a second functional,
\begin{eqnarray}
W[P(N)]  & \equiv & 2 V[P(N)] - V[P(2N] \nonumber \\
& =&  \log[P(2N) \mu / P^2(N)],
\label{eq_WN_def} 
\end{eqnarray}
where by construction this contribution is eliminated
following a suggestion made recently by Semenov and Johner \cite{ANS03}.
The normalization factor $\mu$ appearing in Eq.~(\ref{eq_WN_def}) eliminates again 
a weak vertical scission energy dependence of the data. Obviously, $W[\pchain] \equiv 0$ 
for perfectly Flory distributed chains. 
Following Eq.~(\ref{eq_PNclaim1}) one expects $W[P(N)] = 2 \dmuchain(N) - \dmuchain(2 N)$ and
due to Eq.~(\ref{eq_muEP})
\begin{equation}
\frac{W[\pchain]}{\cep \sqrt{\mu} (2-1/\sqrt{2})} \approx
\frac{1- 0.906 x}{\sqrt{x}}
\label{eq_WN}
\end{equation}
with $x=\mu N$.
Eq.~(\ref{eq_WN}) is indicated by the bold line 
which compares again rather well with the presented data. 

%
%

\begin{figure}[t]
\centerline{\resizebox{0.9\columnwidth}{!}{\includegraphics*{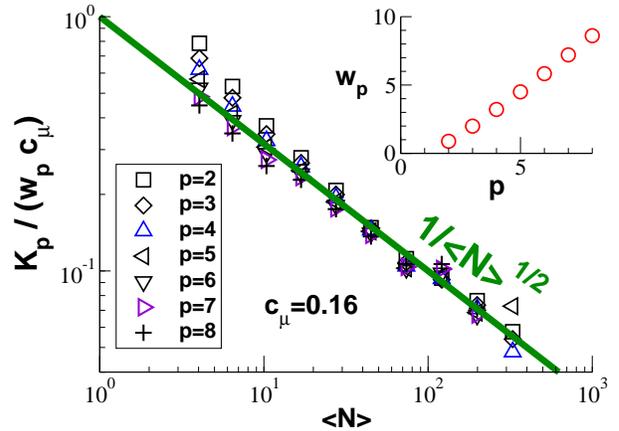}}}
\caption{
Non-exponentiality parameter $\nonexp$ for different moments $p$
as a function of mean chain length $\Nav$. 
$\nonexp$ is finite decreasing with chain length as suggested by Eq.~(\ref{eq_nonexp}).
The vertical axis has been rescaled successfully using the $p$-dependent weights $\wep$ 
indicated in the inset.
%
%
\label{fig_nonexp}
}
\end{figure}

%
%
The functionals presented in Fig.~\ref{fig_WN} require histo\-grams with very high accuracy. 
That $\pchain$ is only approximately Flory distributed can be more readily seen using the
``non-exponentiality parameter" $\nonexp \equiv 1 - \Npav / p! \Nav^p$
which measures how the moments deviate from the Flory distribution.
Obviously, $\nonexp \equiv 0$ for rigorously Gaussian chains.
As stated in Eq.~(\ref{eq_nonexp}), we expect the non-exponentiality parameter
to decay as $\nonexp = \wep \cep \sqrt{\mu} \approx 1/\rho\sqrt{\Nav}$, 
i.e. as the correlation hole potential of the typical melt chain. 
%
The main panel of Fig.~\ref{fig_nonexp} presents 
$\nonexp / \wep \cep$ as a function of $\Nav \approx 1/\mu$
using double-logarithmic axes. The predicted power-law decay is clearly demonstrated by the data.
Note that the scaling of the vertical axis with the $p$-dependent geometrical factors 
$\wep$ allows to bring all moments on the same master curve. 
As can be seen from the inset of Fig.~\ref{fig_nonexp}
this scaling is significant since $\wep$ varies over nearly a decade
between $w_2=\sqrt{\pi}/2$ and $w_8 \approx 8.6$.
Deviations from the predicted scaling are visible, not surprisingly,
for small $\Nav < 10$.
%

\section{Discussion}
\label{sec_disc}

\paragraph*{Summary.}
Challenging Flory's ideality hypothesis, we have investigated in this study the scaling 
of the chemical potential of polymer chains with respect to the length $n$ of a tagged test 
chain plugged into a solution of $N$-chains of a given length distribution $\pchain$ with
$\Nav$ being the typical length of the chains of the bath.
%
%
By means of one-loop perturbation calculations we have demonstrated for $n \ll \Nav^2$
the existence of a non-extensive deviation $\dmuchain(n) \ll 1$ with respect to the Gaussian reference.
This correction becomes universal for small reduced test chain lengths, $x \equiv n/\Nav \ll 1$, scaling as
$\dmuchain(n) \sim 1/\rho \sqrt{n}$ irrespective of the length distribution
as suggested by the ``correlation hole potential" (Fig.~\ref{fig_sketch}(a)).
For larger $x$ the correction depends somewhat on $\pchain$, as explicitly discussed 
for Flory distributed [Eq.~(\ref{eq_muEP})] and monodisperse melts [Eq.~(\ref{eq_mu_MD_Debye})],
but remains generally a monotonously decreasing function of $n$ scaling as
\begin{equation}
\dmuchain(n) \approx \frac{1}{\rho \sqrt{n}} (1-I_c(x))
\mbox{ with }  
I_c(x) \stackrel{x \ll 1}{\Longrightarrow} 0
\label{eq_scaling}
\end{equation}
changing sign at $x \approx 1$ (Fig.~\ref{fig_muchain}).
For the important limit of a test chain of same length as the typical chain of the bath,
$x \equiv 1$, the deviation from Flory's hypothesis decreases in magnitude with $n$.
For Flory distributed or monodisperse chains $I_c(x=1) > 1$ 
and the asymptotic limit, $\muchain(n) \to \mu n$, is thus approached from below 
[Eq.~(\ref{eq_mu_MD_Debye2})].
Note that our predictions are {\em implicit} to the theoretical framework
put forward by Edwards \cite{DoiEdwardsBook} or Sch\"afer \cite{SchaferBook},
but to the best of our knowledge they have not been stated {\em explicitly} before.

We have confirmed theory by analyzing in Section~\ref{sec_simu}
the length distribution of essentially Flory distributed equilibrium polymers obtained for 
different scission energies $E$ by Monte Carlo simulation of the BFM at one melt density.
Albeit the deviations from Flory's hypothesis are small (Fig.~\ref{fig_Florydistr}(a,b)), 
they can be demonstrated by analyzing $-\log(\pchain)$ as shown in Fig.~\ref{fig_WN} 
or from the scaling of the  non-exponentiality parameter, $\nonexp \sim 1/ \sqrt{\Nav}$,
for all moments $p$ sampled (Fig.~\ref{fig_nonexp}). 
We emphasize that the data collapse on the theoretical predictions,
Eqs.~(\ref{eq_VN},\ref{eq_WN},\ref{eq_nonexp}), has been achieved 
without any free adjustable parameter since the coefficient $\cep$ is known.

\paragraph{Outlook.}
Clearly, the presented study begs for a direct numerical verification of the suggested
non-extensive chemical potential for a test chain inserted into a melt
of {\em mono\-disperse} chains, Eq.~(\ref{eq_mu_MD1}).
In principle, this should be feasible by thermodynamic integration using multihistogram 
methods as proposed in \cite{MP94}. In particular, this may allow to improve the numerical
test of the theory for $x \gg 1$; due to the exponential cut-off [Eq.~(\ref{eq_PNclaim})]
this regime has been difficult to explore using the equilibrium polymer length distribution
(Fig.~\ref{fig_WN}).
%
%

\begin{figure}[tb]
\centerline{\resizebox{0.9\columnwidth}{!}{\includegraphics*{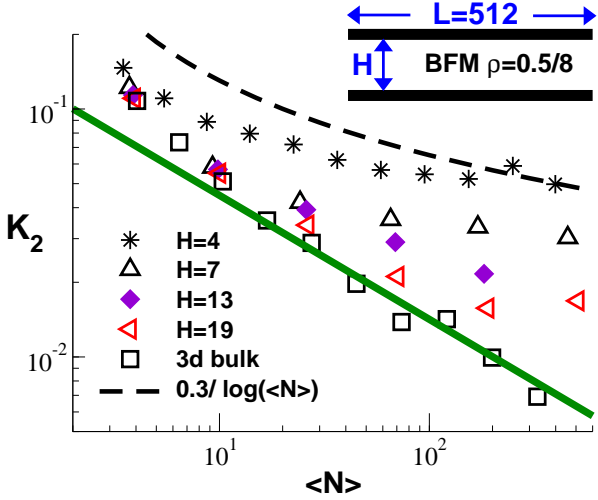}}}
\caption{
Non-exponentiality parameter $K_2=1-\la N^2 \ra /2\Nav^2$ as a function of the mean chain length $\Nav$ 
for equilibrium polymers confined between two hard walls at distance $H$ as indicated.
The lateral box size is $L=512$, the number density $\rho=0.5/8$.
In qualitative agreement with theory \cite{ANS03} 
the non-exponentiality parameter is found to
decay much more slowly with $\Nav$ as in the bulk (squares), i.e. $K_2$ increases with decreasing $H$.
Note that in the large-$\Nav$ limit one expects $K_2 \approx H^0/\log(\Nav)$ 
as indicated by the dashed line. 
%
\label{fig_K2slit}
}
\end{figure}

Another interesting testing bed for the proposed correlation hole effect
are polymer melts confined in thin films of width $H \ll R(n)$ \cite{CMWJB05}. 
A logarithmically decreasing non-extensive chemical potential contribution 
has been predicted for these effectively two-dimensional systems \cite{ANS03,Sergei81}.
The non-exponentiality parameter of equilibrium polymers confined in thin films 
should thus decay rather slowly with chain length. This is in fact confirmed qualitatively 
by the numerical results presented in Fig.~\ref{fig_K2slit} obtained using again the BFM
algorithm with finite scission energy described above. Note that the smallest film width 
allowing the overlap of monomers and the crossing of chains ($H=4$) corresponds to an 
increase of $K_2$ by nearly a decade for the largest chains we have sampled. 
The detailed scaling with $H$ is, however, far from obvious.
Larger mean chain lengths and better statistics are warranted to probe the 
logarithmic behavior for asymptotically long chains predicted by Semenov and Johner \cite{ANS03}.
%
%
Note that if confirmed this prediction should influence the phase diagrams of polymer blends
in reduced effective dimensions.

Finally, we would like to point out that the presented perturbation calculation
for dense polymer chains may also be of relevance to the chemical potential
of dilute polymer chains at and around the $\Theta$-point which has received
attention recently \cite{Rubinstein08}. The reason for this connection is that
(taken apart different prefactors) the {\em same} effective interaction potential,
Eq.~(\ref{eq_veff_q2}), enters the perturbation calculation in the low wavevector limit.
A non-extensive correction $\dmuchain(n) \sim +1/\sqrt{n}$ in three dimensions 
is thus to be expected. 

%
%
\begin{acknowledgement}
We thank  the Universit\'e de Strasbourg, the CNRS, 
and the ESF-STIPOMAT programme for financial support. 
We are indebted to S.P.~Obukhov and A.N.~Semenov for helpful discussions.
\end{acknowledgement}


\begin{thebibliography}{27}
\bibitem{foot_PN}
We suppose throughout this paper that $\pchain$ is a realistic polymer length
  distribution which is not too broad. All moments $\Npav^{1/p}$ exist and are
  of same order. Obviously, all moments of monodisperse melts of length $N$
  become $\Npav^{1/p}=N$.
\bibitem{foot_FloryHuggins}
The chain length distribution is obtained by minimizing a Flory-Huggins free
  energy functional $$f[\rhochain] = \sum_N \rhochain \left( \log(\rhochain) +
  \mu N + E + \dmuchain(N) \right)$$ with respect to the density $\rhochain =
  \rho \pchain/\Nav$ of chains of length $N$. The first term on the right is
  the usual translational entropy. The second term entails a Lagrange
  multiplier which fixes the total monomer density $\rho = \sum_N N \rhochain$.
  All contributions to the chemical potential of the chain $\muchain$ which are
  linear in $N$ can be adsorbed within the Lagrange multiplier. The scission
  energy $E$ characterizes the enthalpic free energy cost for breaking a chain
  bond. The most crucial last term encodes the remaining non-linear
  contribution $\dmuchain(N)$ to the chemical potential $\muchain$ which has to
  be paid for creating two new chain ends. A rigorously Flory distributed
  length distribution implies thus $\dmuchain(N) = const$.
\end{thebibliography}

\end{document}